\documentclass[mathleft]{an}
\usepackage{times,graphicx,epsfig}
\newcommand{\bq}{\begin{equation}}
\newcommand{\eq}{\end{equation}}
\newcommand{\bqn}{\begin{eqnarray}}
\newcommand{\eqn}{\end{eqnarray}}

\begin{document}
\title{The main sequence from F to K stars of the solar neighbourhood in SDSS colours}
\author{Andreas Just\thanks{Corresponding author:
  \email{just@ari.uni-heidelberg.de}}
\and        Hartmut Jahrei\ss
}
\titlerunning{The main sequence of the solar neighbourhood 
in SDSS colours}
\authorrunning{A. Just, H. Jahrei\ss}
\institute{Astronomisches Rechen-Institut at ZAH, University of
Heidelberg, M\"onchhofstra{\ss}e 12-14, 69120
Heidelberg, Germany
}

\received{XXXX}
\accepted{XXXX}
\publonline{XXXX}
\keywords{(Galaxy:) solar neighborhood, Galaxy: stellar content,
stars: fundamental parameters, catalogs}

\abstract{For an understanding of Galactic stellar populations in the SDSS 
filter system well defined stellar samples are needed. 
The nearby stars provide a complete
stellar sample representative for the thin disc population. 
We compare the filter transformations of different authors applied to
the main sequence stars from F to K dwarfs to SDSS filter system 
and discuss the properties of the main sequence. 
The location of the mean main sequence in colour-magnitude diagrams is very
sensitive to systematic differences in the filter transformation.
A comparison with fiducial sequences of star clusters observed in g',r',i' show
good agreement. Theoretical isochrones from Padua and from Dartmouth have still
some problems especially in (r-i)-colour.}

\maketitle

\section{Introduction\label{sec-introd}}

The Sloan Digital Sky Survey (SDSS) with  sky coverage of almost 10,000 deg$^2$
has collected at present the largest and most homogeneous database comprising 
about $10^8$ stellar objects in the Milky Way. The SDSS photometric system was 
specifically designed for the survey to provide continuous coverage over the 
entire optical wavelength range in the five filters $u\, g\, r\, i\, z$. 
In the most recent sixth data release
(Adelman-McCarthy et al. \cite{Ad08}) the new 'ubercalibration' of the photometry was
applied to provide an improved homogeneous relative calibration over the whole 
survey to
1\% in $g\, r\, i\, z$ and 2\% in $u$ (Padmanabhan et al. \cite{Pa08}). The
zero-points are not exactly in the AB system, but for the $g\, r\, i$ filters,
the corrections are below 1\% and can be neglected. For the zero-points of stars
brighter than 14.5\,mag see also Chonis \& Gascell (\cite{Ch08}).

Since most standard stars are too bright to be directly observed with the 
2.5\,m SDSS telescope, 
it is a complicated matter to establish standards.
One way is to measure the standard stars with a smaller telescope.
The SDSS photometric system was 
originally defined at the 0.5\,m photometric telescope (PT) with
$u'\, g'\, r'\, i'\, z'$ (Fukugita et al. \cite{Fu96}). 
A similar filter set is used in the USNO
1\,m telescope for these calibration issues. Since these filters are slightly
different from $u\, g\, r\, i\, z$ at the main 2.5\,m telescope 
(Abazajian et al. \cite{Ab03}), transformation
formulas from $u'\, g'\, r'\, i'\, z'$ to $u\, g\, r\, i\, z$ must be 
established (Tucker et al. \cite{Tu06}). 

Despite the enormous wealth of information that the SDSS database provides the
majority of current observational and theoretical knowledge of resolved stellar
populations is based largely on the Johnson-Kron-Cousins $U\,B\,V\,R_\mathrm{C}\,I_\mathrm{C}$ 
photometric
system and some other systems such as the Str\"omgren, DDO, Vilnius, and Geneva
systems. 

The best way of calibrating the theoretical isochrones is the observation of 
fiducial isochrones of simple stellar populations (SSP) with known age and
metallicity in SDSS filters.
Clem, VandenBerg, \& Stetson (\cite{Cl08}) obtained fiducial 
stellar population sequences for the $u'\, g'\, r'\, i'\, z'$ photometric 
system from a number of different Galactic star clusters, spanning a wide range 
in metallicity and magnitude. In order to achieve the required accuracy they 
have first established a new set of secondary standard stars in selected 
open and globular star clusters for this photometric system 
(Clem, VandenBerg, \& Stetson \cite{Cl07}).

For a physical understanding of observed stellar populations is essential to
interpret the data via stellar evolution tracks or isochrones.
Girardi et al. (\cite{Gi04}) were the first, who provided tables of 
theoretical isochrones in $u\, g\, r\, i\, z$ derived from stellar spectra. 
These have however turned out to lack the necessary precision to correctly 
interpret different stellar populations from the SDSS database. 
Recently in the Dartmouth Stellar Evolution Program 
(Dotter et al. \cite{Do07,Do08})
isochrones in $u\, g\, r\, i\, z$ are presented based on VandenBerg \& Clem
(\cite{Va03}) and Ivezi\'c et al. (\cite{Iv07}).

The solar neighbourhood provides a complete stellar sample of the Milky
Way disc with measured absolute magnitudes from Hipparcos data and the 
Catalogue of Nearby Stars (CNS4, Jahreiss \& Wielen \cite{jah97}). 
Therefore it is a unique place for
investigating the properties of the stellar disc population testing the
transformations to the SDSS filters.
The purpose of this paper is to transform the CNS4 
absolute luminosities in 
($M_\mathrm{B}\,M_\mathrm{V}\,M_{\mathrm{R_C}}\,M_{\mathrm{I_C}}$) of main sequence (MS)
stars
to the SDSS luminosities ($M_\mathrm{g}\, M_\mathrm{r}\, M_\mathrm{i}$), and to derive the corresponding
colour-magnitude relations.

In the next section we present the filter transformations of different authors.
In Sect. \ref{sec-stars} the selection of the stellar sample and the available
colours are presented.
In Sect. \ref{sec-trafo} we discuss the differences of the various
transformation formulas.
In Sect. \ref{sec-ms} we derive the MS properties for F-K stars and
discuss the uncertainties due to different transformations. A comparison
with observed fiducial isochrones is also presented and the local stellar number
density in $g-r$ space is derived.
In Sect. \ref{sec-iso} a comparison
with theoretical isochrones in
the colour magnitude diagrams (CMD) is given.
The results are summarized in Sect. \ref{sec-disc}.

\section{Filter transformations\label{sec-filter}}

Several groups have derived photometric transformations of the SDSS 
photometric system to the $U\,B\,V\,R_\mathrm{C}\,I_\mathrm{C}$ system or vice versa. 
Smith et al. (\cite{Sm02}) defined the $u'\, g'\, r'\, i'\, z'$ photometric 
system on 158 standard stars, a subset of $U\,B\,V\,R_\mathrm{C}\,I_\mathrm{C}$ standard stars 
from Landolt (\cite{La92}), using
the USNO-1.0m telescope. They derived transformation formulas from 
$U\,B\,V\,R_\mathrm{C}\,I_\mathrm{C}$ to
$u'\, g'\, r'\, i'\, z'$. Rodgers et al. (\cite{Ro06}) refined this work
by including only MS stars into the transformation calculation
and using a second-order polynomial in 
($B-V$) for $g'$ as well as bilinear equations including two colours, where the
 corresponding
filters overlap, i.e. $B-V$ and $V-R_\mathrm{C}$ for calculating $g'-r'$. 
Bilir, Karaali, \& Tun\c{c}el (\cite{Bi05}) also improved the transformations of 
Smith et al. (\cite{Sm02}) by restricting the stellar sample to population I
stars and checking the stellar locus in the 2-colour diagram (g-r,r-i) by
comparing with a sample of nearby Hipparcos stars.  

Jordi, Grebel, \& Ammon (\cite{Jo06}) have derived the colour transformation for main 
sequence stars 
directly for the 2.5m SDSS telescope $u\, g\, r\, i\, z$ photometric system 
by including the faint extension of standard stars from Stetson (\cite{St00}).
Separate transformation formulas were derived for population I and II stars.
Chonis \& Gascell (\cite{Ch08}) derived linear transformations  from 
$g\, r\, i$ to $B\,V\,R_\mathrm{C}\,I_\mathrm{C}$ using the Landolt and Stetson standard stars,
which are inverted for our purposes. They also showed the consistency with 
the transformations of Jordi et al. (\cite{Jo06}).

Ivezi\'c et al. (\cite{Iv07}) derived nonlinear functions for the
transformations from $g\, r\, i$ to $B\,V\,R_\mathrm{C}\,I_\mathrm{C}$.
Recently Bilir et al. (\cite{Bi08}) have calculated transformations between 
the SDSS photometry and the
2MASS photometry as well as between the $U\,B\,V\,R_\mathrm{C}\,I_\mathrm{C}$ photometry and the
2MASS photometry for different metallicity ranges.

We compare the results of four different transformations, namely
of Bilir et al. (\cite{Bi05}) labeled by B05,
of Jordi et al. (\cite{Jo06}) J06,
of Rodgers et al. (\cite{Ro06}) R06,
and of Chonis \& Gaskell (\cite{Ch08}) C08.
Since B05 and R06 are both improvements of the transformations of 
Smith et al. (\cite{Sm02}) labeled by S02, we do not include S02 in the
investigation.
The transformations of Ivezi\'c et al. (\cite{Iv07}) are not included, because
they cannot easily be inverted due to the higher-order terms. 
Applying the transformations of Bilir et al. (\cite{Bi08}) via the 2MASS 
photometry without taking 2MASS data of the stellar sample into account seems
also not reasonable.

We cannot claim at the present status of our investigations, which of the
transformations should be preferred.
The transformation formulas for $M_\mathrm{g}$ and $g-r$ of B05 and C08
can be applied without further approximations to the complete stellar sample,
because only $M_\mathrm{V}$ and $B-V$ are required. 
For convenience and clarity in the 
comparisons of the different transformations
we select the C08 transformation as fiducial.

In J06 we use the transformations for population I stars,
where we combined the transformations for $g-V$ and $r-R$ to derive 
an equation for $g-r$ as a function of $B-V$ and $V-R_\mathrm{C}$
instead of using the simple equation for $g-r$ given in their table. 
The stellar locus in the 2-colour diagram ($B-V,V-R_\mathrm{C}$), where these equations
are consistent to each other, is shown in Fig. \ref{fig-2col}.

The transformations R06 are originally to the primed system and must be
converted to the unprimed system.

Throughout the paper we
convert all colours from the $u'\, g'\, r'\, i'\, z'$ to the 
$u\, g\, r\, i\, z$ system, if necessary, by applying
\bqn
g &=& g'+0.060[ (g'-r')-0.53 ] \nonumber\\
r &=& r'+0.035[ (r'-i')-0.21 ]  \label{eq-griTu}\\
i &=& i'+0.041[ (r'-i')-0.21 ] \nonumber
\eqn
from  Tucker et al. (\cite{Tu06}) leading to
\bqn
g-r &=& 1.060(g'-r')-0.035(r'-i')-0.024  \label{eq-grTu}\\
r-i &=& 0.994(r'-i')+0.001 \nonumber
\eqn
Note, that transformations via the primed system
lead to additional colour terms in the unprimed system, i.e. ($V-R_\mathrm{C}$)-terms
for $g$ and $g-r$ and also a ($R_\mathrm{C}-I_\mathrm{C}$)-term in the formula for $g-r$.

All transformations can be written in the general form
\bqn
g -V&=& a_1(B-V)^2 + b_1(B-V)+ c_1(V-R_\mathrm{C}) +d_1 \nonumber
\eqn
\bqn
g-r &=&a_2(B-V) + b_2(V-R_\mathrm{C}) +c_2(R_\mathrm{C}-I_\mathrm{C})+d_2  \nonumber \\
r-i &=& a_3(R_\mathrm{C}-I_\mathrm{C}) +b_3\label{eq-gri}
\eqn
The coefficients of B05, J06, R06, and C08 are listed in Table \ref{tab-coeff}
with the coefficients of S02 (Smith et al. \cite{Sm02}) added for comparison.
\begin{table*}
\caption {Coefficients of the transformation formulas given in Eqs. \ref{eq-gri}
of the different authors. The first 3 rows refer to the calculated quantity, the
contributing terms and the coefficients, respectively. 
Column 1 gives the source, and the other columns the coefficient, if the terms
appear in the corresponding equation. In the notes the validity ranges are also
given.
}
\begin{center}
\begin{tabular*}{18cm}{l@{\extracolsep\fill}|cccc|cccc|cc}
\hline 
formula& \multicolumn{4}{l|}{$M_\mathrm{g}-M_\mathrm{V}$}& \multicolumn{4}{l|}{g-r}& \multicolumn{2}{l}{r-i}\\
term & $(B-V)^2$& $(B-V)$& $(V-R_\mathrm{C})$&cnst.& $(B-V)$& $(V-R_\mathrm{C})$&$(R_\mathrm{C}-I_\mathrm{C})$&cnst.& $(R_\mathrm{C}-I_\mathrm{C})$&cnst.\\
coeff. & $a_1$& $b_1$& $c_1$& $d_1$& $a_2$& $b_2$& $c_2$& $d_2$& $a_3$& $b_3$ \\
\hline 
C08 & -- & 0.642 & -- & $-0.135$ & 1.094 & -- & -- & $-0.248$ & 0.939 & $-0.198$\\
R06 & $-0.042$ & 0.619 & 0.079 & $-0.133$ & 0.295 & 1.365 & $-0.035$ & $-0.249$ & 0.994 & $-0.210$ \\
J06 & -- & 0.643 & -- & $-0.127$ & 0.643 & 0.725 & -- & $-0.213$ & 0.988 & $-0.221$ \\
B05 & -- & 0.634 & -- & $-0.108$ & 1.124 & -- & -- & $-0.252$ & 1.04 & $-0.224$ \\
S02 & -- & 0.60 & -- & $-0.11$ & 1.039 & -- & $-0.035$ & $-0.218$ & 0.994 & $-0.21$ \\
\hline
\end{tabular*}
\end{center}
C08: Chonis \& Gascall (\cite{Ch08}): $0.08<r-i<0.5$ and $0.2<g-r<1.4$ \\
R06: Rodgers et al. (\cite{Ro06}): $-0.2<B-V<1.6$ \\
J06: Jordi et al. (\cite{Jo06}): $g-r$ for $(V-R_\mathrm{C}) \le 0.93$ \\
B05: Bilir et al. (\cite{Bi05}): $0<B-V<1.4$ \\
S02: Smith et al. (\cite{Sm02}):  $-0.3<B-V<1.9$ 
	(and $R_\mathrm{C}-I_\mathrm{C} \le 1.15$ for $r'-i'$)
\label{tab-coeff}
\end{table*}

\section{Nearby stars\label{sec-stars}}

\begin{figure*}
\includegraphics[width=8.5cm, angle=0]{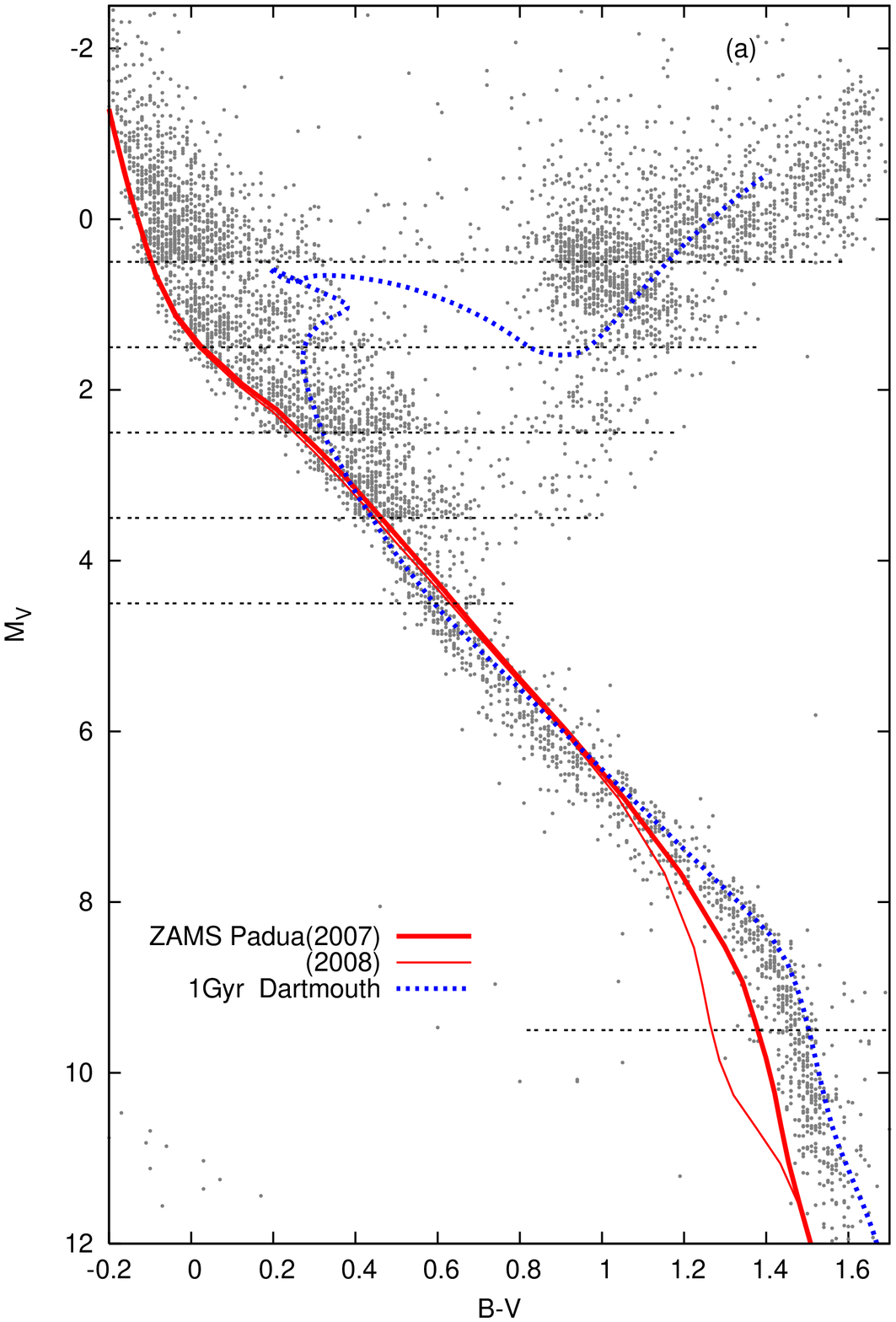}
\includegraphics[width=8.5cm, angle=0]{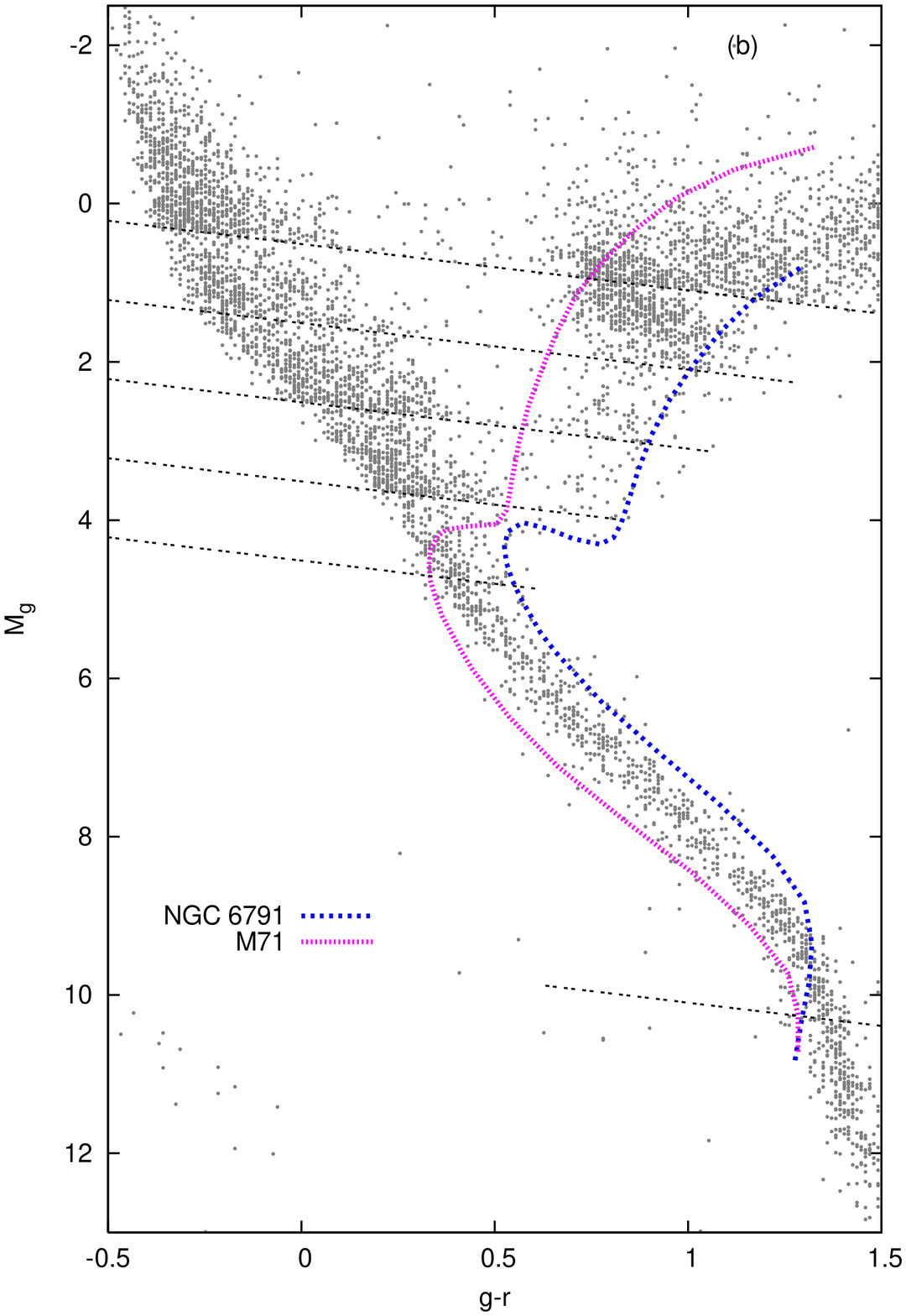}
\caption{ CMDs in (B-V, M$_\mathrm{V}$) and (g-r, M$_\mathrm{g}$), where the transformation C08
 was used (Chonis \& Gascall \cite{Ch08}), for volume-selected
Hipparcos stars (grey dots; see Sect.\ \ref{sec-stars}). The completeness distances are 200, 100,
75, 50, 30, 25\,pc in the magnitude ranges $<0.5$, 1$\pm0.5$, 2$\pm0.5$, 
3$\pm0.5$, 4$\pm0.5$, and 7$\pm0.5$\,mag,
respectively. In (a) the ZAMS from Padua (old: thick line, new: thin line)
 and the 1\,Gyr isochrone from Dartmouth
with metallicity [Fe/H]=+0.13 (see Sect.\ \ref{sec-iso}) are overplotted.
In (b) the ridge lines of the star clusters NGC 6791 and M71 are shown.
\label{fig-hrd}}
\end{figure*}

We start with a volume complete sample of nearby stars selected from Hipparcos stars and
the CNS4 to the completeness limit in magnitude bins of $M_\mathrm{V}$. 
Due to the Hipparcos survey the limiting distances are 200, 100,
75, 50, 30, 25\,pc in the magnitude ranges $<0.5$, 1$\pm 0.5$, 2$\pm 0.5$, 
3$\pm 0.5$, 4$\pm 0.5$, and 7$\pm 2.5$\,mag,
respectively. The CMDs in Hipparcos colours ($B-V$, $M_\mathrm{V}$) and in
($g-r$, $M_\mathrm{g}$) using C08 are shown in Fig.\ \ref{fig-hrd}. Here fainter stars
are included in order to show the turnover of the MS to the M dwarfs. The
transformations C08 are valid only for $0.4<B-V<1.5$. The extrapolation to the
blue and red part as well as to the giants must be used with caution. 

For the detailed investigation of the local MS from F to K dwarfs we
use the data from the Catalogue of Nearby Stars (CNS4). All stars are also in
 the
Hipparcos catalogue, but the luminosities and colours given in the CNS4 rely on
more thoroughly collected photometric data than from  
the Hipparcos data alone.

F to K stars were selected from the CNS4 
according to  $ 0.30 \le B-V \le 1.35 $ with parallaxes larger than 40\,mas. 
We removed stars more than 0.8 mag above and below the 
estimate mean MS in order to avoid contamination by turnoff stars, giants and
white dwarfs. After removing additionally a few stars with poor photometry 
there remained 786 dwarf stars. 
We use this stellar sample as a complete sample of MS stars from F to K type
inside a sphere with 25\,pc radius.
$M_\mathrm{V}$,   $B-V$, $V-R$, $R-I$ from the CNS4 catalogue and
$g-r$,  $r-i$,  $M_\mathrm{g}$ from the C08 transformation is given as online material
for these stars. 
The first few lines are shown in Table \ref{tab-data}.
\begin{table*}
\caption {First lines of the table with luminosities and colours of the 
complete sample of 786 stars in the 25\,pc sphere. 
Column 1 is the Hipparcos number, 
column 2--4 are $g-r$,  $r-i$,  $M_\mathrm{g}$ from the C08 transformation,
column 5--8 are $M_\mathrm{V}$,   $B-V$, $V-R$, $R-I$ from the CNS4 catalogue, 
and column 9 is the CNS identifier; suffix A and B denotes components of
binaries. 
The full table is available as online material with '|' as separators.
}
\begin{center}
\begin{tabular*}{11cm}{r|c|c|c|c|c|c|c|l}
\hline 
  Hip  &  $g-r$  &  $r-i$  &  $M_\mathrm{g}$  &   $M_\mathrm{V}$  &   $B-V$  & $V-R$   & $R-I$   &  CNS\\
\hline 
   sun & 0.459 & 0.114 & 5.110 &   4.83 & +0.646 & 0.354 & 0.332 & 1    \\ 
   171 & 0.484 & 0.186 & 5.684 &   5.39 & +0.669 & 0.409 & 0.409 & 19140A\\
   436 & 0.915 & 0.320 & 8.017 &   7.47 & +1.063 & 0.636 & 0.552 & 1  30 \\
   518 & 0.501 &       & 5.135 &   4.83 & +0.685 & 0.380 &       & 1  41A\\
   544 & 0.575 &       & 5.788 &   5.44 & +0.752 & 0.410 &       & 1  50 \\
   910 & 0.267 & 0.074 & 3.637 &   3.47 & +0.471 & 0.305 & 0.290 & 1 100 \\
   950 & 0.238 &       & 3.710 &   3.56 & +0.444 &       &       & 4  15 \\
  1031 & 0.600 &       & 6.043 &   5.68 & +0.775 &       &       & 22001 \\
  ...
\end{tabular*}
\end{center}
\label{tab-data}
\end{table*}

From these 786 stars  
561 stars have not only a $B-V$ colour but also $V-R_\mathrm{C}$ and $R_\mathrm{C}-I_\mathrm{C}$ colours.
We use this subsample as representative for the locus of the MS in
the solar neighbourhood.
For 119 stars $R_\mathrm{C}-I_\mathrm{C}$ is missing, for 6 stars $V-R_\mathrm{C}$ is  missing, 
and 100 stars both colours are missing.

\section{Transformation differences}\label{sec-trafo}

We use the subsample of 561 stars with $M_\mathrm{V}$, $B-V$, $V-R_\mathrm{C}$, $R_\mathrm{C}-I_\mathrm{C}$
as representative for the properties of the MS in
the solar neighbourhood. 
In order to measure the sensitivity of the CMDs on the transformation formulas
we apply all transformations C08, R06, J06, and B05
(Chonis \& Gascall \cite{Ch08}, Rodgers et al. \cite{Ro06},   
Jordi et al. \cite{Jo06}, Bilir et al. \cite{Bi05}) collected
in Table \ref{tab-coeff} to the same stellar sample and compare the results. 
\begin{figure}
\includegraphics[width=8.5cm, angle=0]{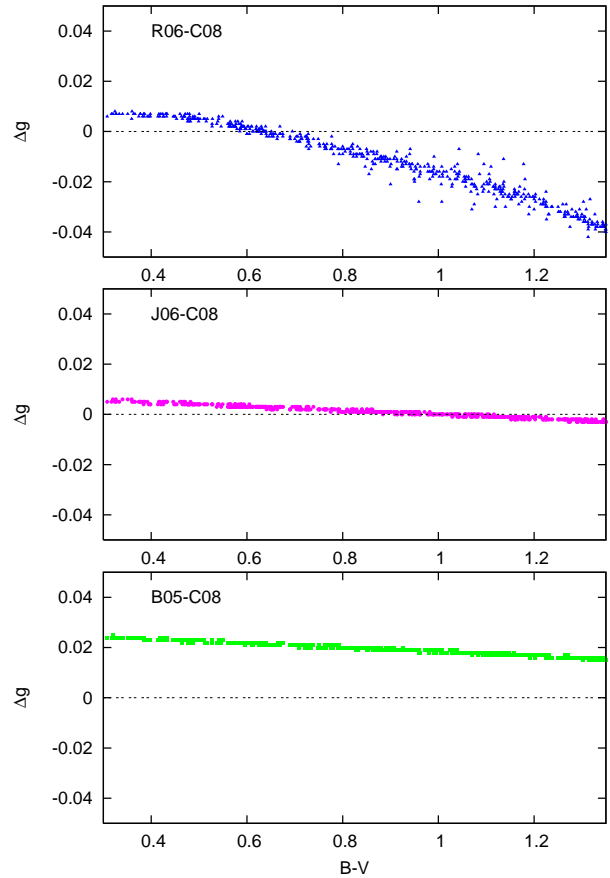}
\caption{ Comparison of the different transformations relative to
C08. The difference in $M_\mathrm{g}$ magnitude
 for the 561 F- to K-dwarfs with all colours in B,V,R,I of the initially selected  
786 stars is plotted as a function of $B-V$.
\label{fig-dg}
}
\end{figure} 
\begin{figure}
\includegraphics[width=8.5cm, angle=0]{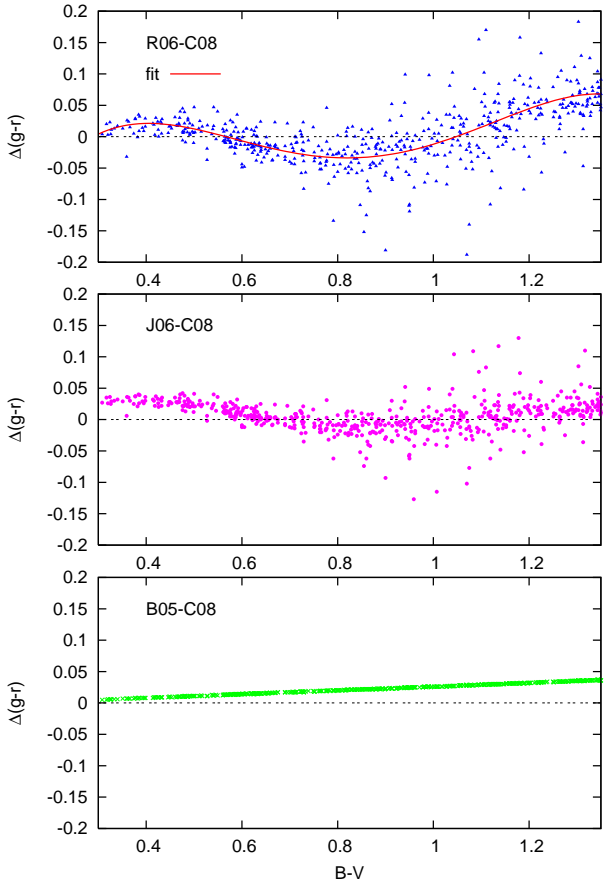}
\caption{ Same as in Fig. \ref{fig-dg} but for $g-r$. 
The full line in the upper plot is
the best fit used for deriving the local number density as a function of $g-r$.
\label{fig-dgr}
}
\end{figure} 
\begin{figure}
\includegraphics[width=8.5cm, angle=0]{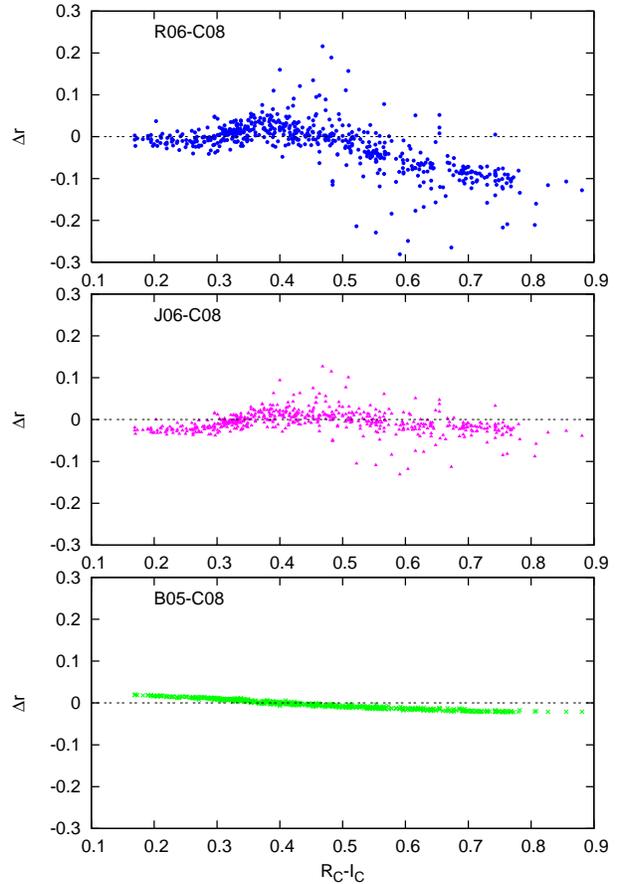}
\caption{ Same as in Fig. \ref{fig-dg}, but for $M_\mathrm{r}$.
\label{fig-dr}
}
\end{figure} 
\begin{figure}
\includegraphics[height=8.5cm, angle=270]{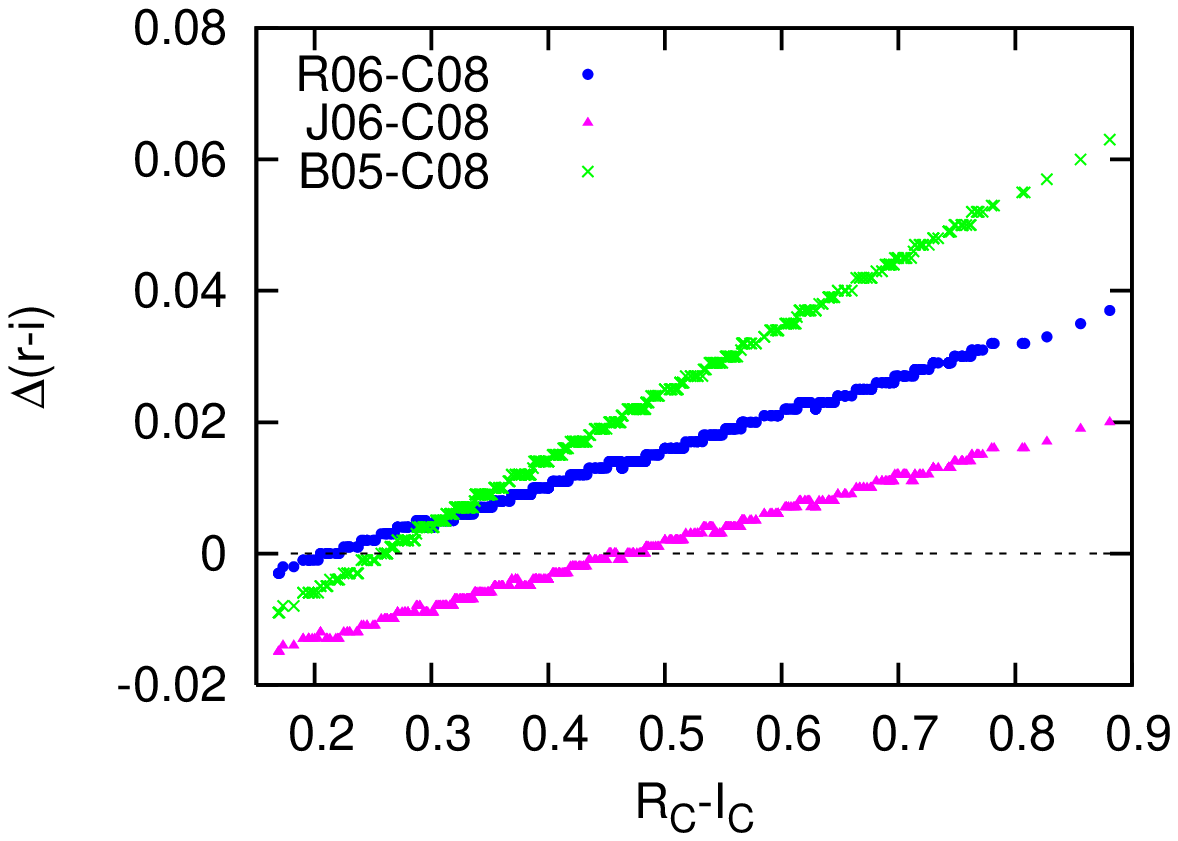}
\caption{ Same as in Fig. \ref{fig-dgr} but for $r-i$.
\label{fig-dri}
}
\end{figure} 

We use the transformations C08 as reference. In Figs. \ref{fig-dg} -
\ref{fig-dri} the differences of the transformations are plotted. The differences
between the other pairs of transformations can be deduced from the differences of
the values in the two corresponding plots with respect to C08.

In $M_\mathrm{g}$ (Fig. \ref{fig-dg}) the systematic differences along the MS are below
0.06\,mag. The scatter in the plot for R06-C08 is exclusively due to the
transformation from the primed to the unprimed filter system.
The differences in the colour index $g-r$ (Fig. \ref{fig-dgr}) 
show a systematic variation of the same order as for $\Delta g$, but with a
different $B-V$ dependence. For the R06 and J06 transformations the scatter
arises from the $V-R_\mathrm{C}$ dependence of $g-r$ and is 
much larger (up to 0.3\,mag).
In $M_\mathrm{r}$ (Fig. \ref{fig-dr}) the variation and scatter is comparable to that in
$g-r$.
In $r-i$ (Fig. \ref{fig-dri}) there is only a linear trend up to 0.08\,mag due
to the simple transformation equations in all cases. 

In total the variations in colour by the different transformations will have a
much stronger influence on the location of the MS than 
the variations in the luminosities.

\section{The main sequence for F to K dwarfs}\label{sec-ms}

The position of the MS in the CMDs and in the 2-colour diagrams are important for
estimating the distance of a stellar population and their properties like
age and metallicity. 
The MS in the solar neighbourhood shows a considerable width due to the spread
in age and metallicity and the contamination by binary stars. 
No clear maximum in the magnitude bins exists (ridge line). 
Therefore we determine a mean MS and measure the width in the CMDs and in the
2-colour diagrams.

The comparison with independent determinations of the MS are
necessary for judging the quality of the different transformations. We use the
fiducial stellar sequences in $g'\, r'\, i'$ of two star clusters from 
Clem et al. (\cite{Cl07,Cl08}) transformed to $g\, r\, i$. 
The old open star cluster NGC 6791 has a metallicity of [Fe/H]=+0.40 
and the reddening is E(B-V)=0.155 (Clem et al. \cite{Cl07}). The age of
8\,Gyr and distance modulus of $m-M = 13.17$ are taken from Carraro et al. 
(\cite{Ca06}). 
For the globular cluster M~71 with metallicity [Fe/H]=-0.58 reddening
and distance modulus are still under debate. 
We use $m-M = 12.89$ and $E(B-V) = 0.25$ derived from $E(J-K) = 0.13$ of
Kyeong, Byun, \& Chun (\cite{Ky97}) using NIR photometry.
These star clusters bracket the metallicity range of
disc stars in the solar neighbourhood. 
Therefore the ridge lines of these simple stellar 
populations (SSP) should be comparable to the boundaries of the local MS.

Fig.\ \ref{fig-hrd} shows a general consistency of the C08 transformation with
the ridge lines of NGC 6791 and M~71 for F to K stars. In the regime of M dwarfs
there appears a discrepancy. But here we are outside the valid colour regime of
C08 and the star cluster fiducial end.

\subsection{Colour-magnitude diagrams}\label{sec-cmd}

The mean absolute magnitudes $M_\mathrm{g}$ as a function of 
$g-r$ is determined by LOWESS (Cleveland \cite{Cl81}) carrying out a robust locally 
weighted regression, where the amount of smoothing is determined by a parameter f 
(in our case $f=0.2$), which describes the fraction of points used to compute each fitted value. 
The result of C08 is presented in 
Table \ref{tab-ms}. 
\begin{table}
\caption {Mean MS $M_\mathrm{g}(g-r)$ in 0.05\,mag bins with the C08 transformation 
for the complete sample of 786 stars in the 25\,pc sphere. 
First column is $g-r$, second column mean $M_\mathrm{g}$, third column the standard
deviation in $M_\mathrm{g}$ and column four the number of stars in the 0.05\,mag bins.
}
\begin{center}
\begin{tabular}{cccc||cccc}
$g-r$& $M_\mathrm{g}$ & $\sigma_\mathrm{g}$ & $N$ &
$g-r$& $M_\mathrm{g}$ & $\sigma_\mathrm{g}$ & $N$ \\
 0.10  &  2.72  &    0.23  &    11   &   0.70  & 6.49   &    0.32  &    45   \\
 0.15  &  3.05  &    0.30  &	14   &   0.75  & 6.72	&    0.29  &	41   \\
 0.20  &  3.39  &    0.26  &	17   &   0.80  & 6.93	&    0.24  &	40   \\
 0.25  &  3.73  &    0.30  &	25   &   0.85  & 7.16	&    0.25  &	32   \\
 0.30  &  4.12  &    0.39  &	33   &   0.90  & 7.39	&    0.29  &	35   \\
 0.35  &  4.49  &    0.39  &	31   &   0.95  & 7.62	&    0.27  &	29   \\
 0.40  &  4.82  &    0.30  &	53   &   1.00  & 7.85	&    0.29  &	41   \\
 0.45  &  5.09  &    0.30  &	58   &   1.05  & 8.08	&    0.20  &	30   \\
 0.50  &  5.38  &    0.30  &	32   &   1.10  & 8.32	&    0.22  &	21   \\
 0.55  &  5.69  &    0.36  &	39   &   1.15  & 8.55	&    0.23  &	28   \\
 0.60  &  5.97  &    0.26  &	36   &   1.20  & 8.80	&    0.27  &	38   \\
 0.65  &  6.24  &    0.34  &	41   &
\end{tabular}
\end{center}
\label{tab-ms}
\end{table}

\begin{figure}
\includegraphics[width=8.5cm, angle=0]{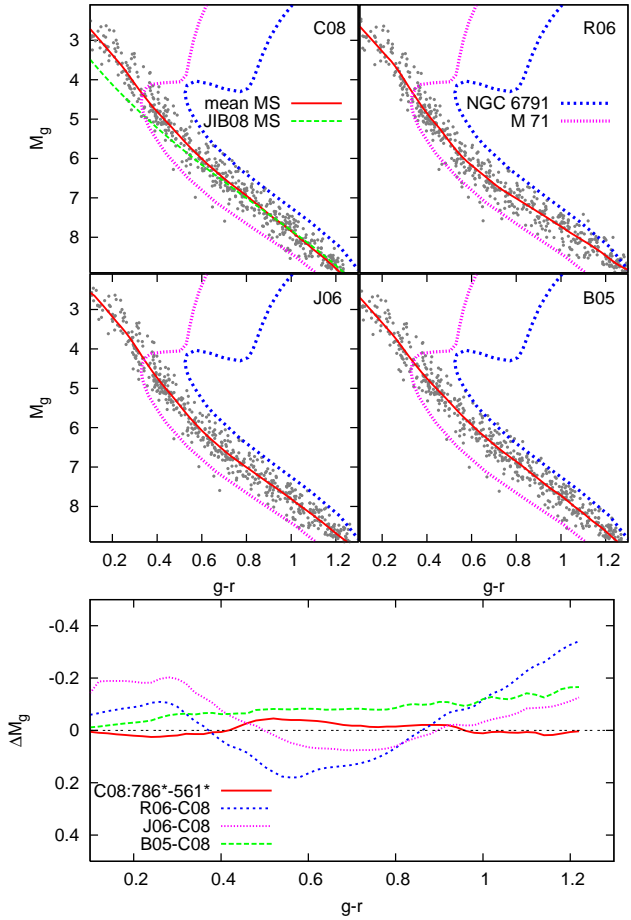}
\caption{CMDs for 561 F- to K-dwarfs within 25\,pc in 
($g-r,M_\mathrm{g}$).
Upper panels: CMDs using the different transformations as noted.
The fiducial colour magnitude sequences for 
NGC~6791 and M71 are shown for comparison.
The full lines give
the mean MS derived with the smoothing algorithm.
For comparison the MS with the bright normalization (JIB08) used in 
Juri\'{c} et al. (\cite{Ju08}) is plotted in the upper left panel.
Lower panel: Differences of the mean MS derived with the different
transformation formulas with respect to C08 (561 stars). 
The difference of the
C08 transformation using the complete sample of 786 stars is also shown.
\label{fig-ms}
}
\end{figure}

\begin{figure}
\includegraphics[width=8.5cm, angle=0]{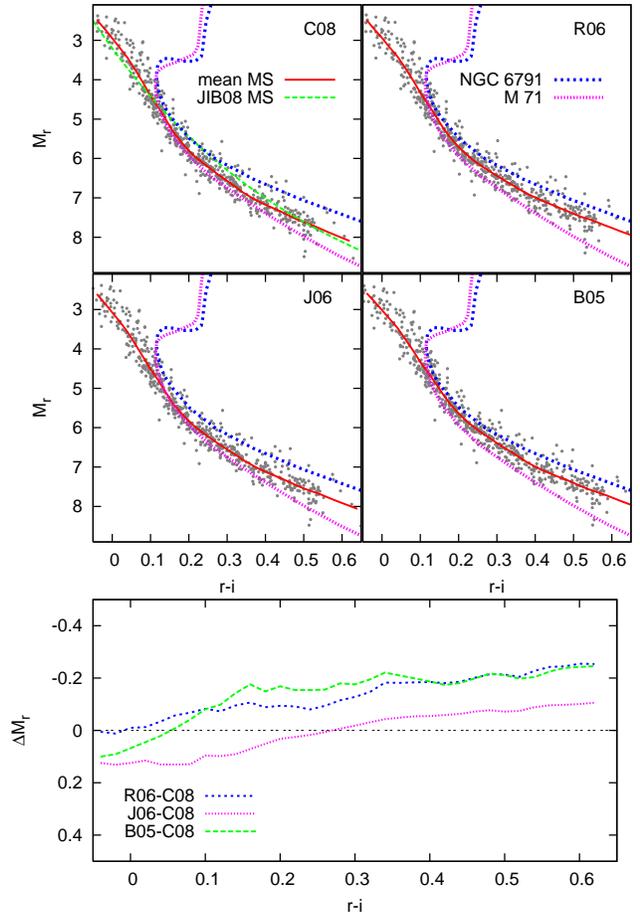}
\caption{Similar plots as in Fig.\ \ref{fig-ms} but in ($r-i,M_\mathrm{r}$).
\label{fig-ms2}
}
\end{figure}

The results for all transformations are shown in Fig. \ref{fig-ms} for 
($g-r,M_\mathrm{g}$) and  in Fig. \ref{fig-ms2}
for ($r-i,M_\mathrm{r}$). 
Dots are the 561 stars 
and the full line is the mean MS.
Overplotted are the fiducial colour magnitude sequences for the star
clusters M71 and NGC~6791 from Clem \& VandenBerg (\cite{Cl08}) 
with metallicities $\mathrm{[Fe/H]}=-0.58$ and +0.40, respectively. 

The upper left panel of Fig. \ref{fig-ms} shows the C08 transformation. 
In ($g-r,M_\mathrm{g}$) the ridge lines of M71 and NGC~6791 bound nicely the local MS
stars.  
The other three panels show the same using the transformations R06, J06, and
B05, respectively.
There is also  a general agreement with the star cluster fiducial. 
Only for R06 there are some outliers at the red and metal rich end.

In the lower panel the mean main
sequences from the different transformations are compared. 
The difference of the
C08 transformation using the complete sample of 786 stars to that with 561 stars
show the uncertainty due to incompleteness. 
The differences are below 0.05\,mag.
The deviations from different transformations are up to 0.5\,mag. 
This corresponds to systematic distance errors of 25\%, when applying the
distance modulus to observed apparent magnitudes. 

In ($r-i,M_\mathrm{r}$) (Fig. \ref{fig-ms2}) there is a clear disagreement above the MS
for all transformations. The trends of the mean main
sequences are smaller than in ($g-r,M_\mathrm{g}$).

The dashed lines in the upper left panels of Figs. \ref{fig-ms} and \ref{fig-ms2}
show the MS with the bright normalization 
used in Juri\'{c} et al. (\cite{Ju08}, JIB08) for comparison. 
In ($r-i,M_\mathrm{r}$), where the MS is originally defined, it fits the data better 
than in ($g-r,M_\mathrm{g}$). 
The locus confirms that the 'bright' MS
represents the local stellar population with mean metallicity 
$\mathrm{[Fe/H]}\approx-0.2$. The offset to the
metal poor regime for F stars is expected, because metal poor F stars have a
lifetime considerably shorter than the age of the stellar disc. Therefore these
old F stars are already in the turnoff phase.


\subsection{2-colour diagrams}\label{sec-2col}

\begin{figure}
\includegraphics[width=8.5cm, angle=0]{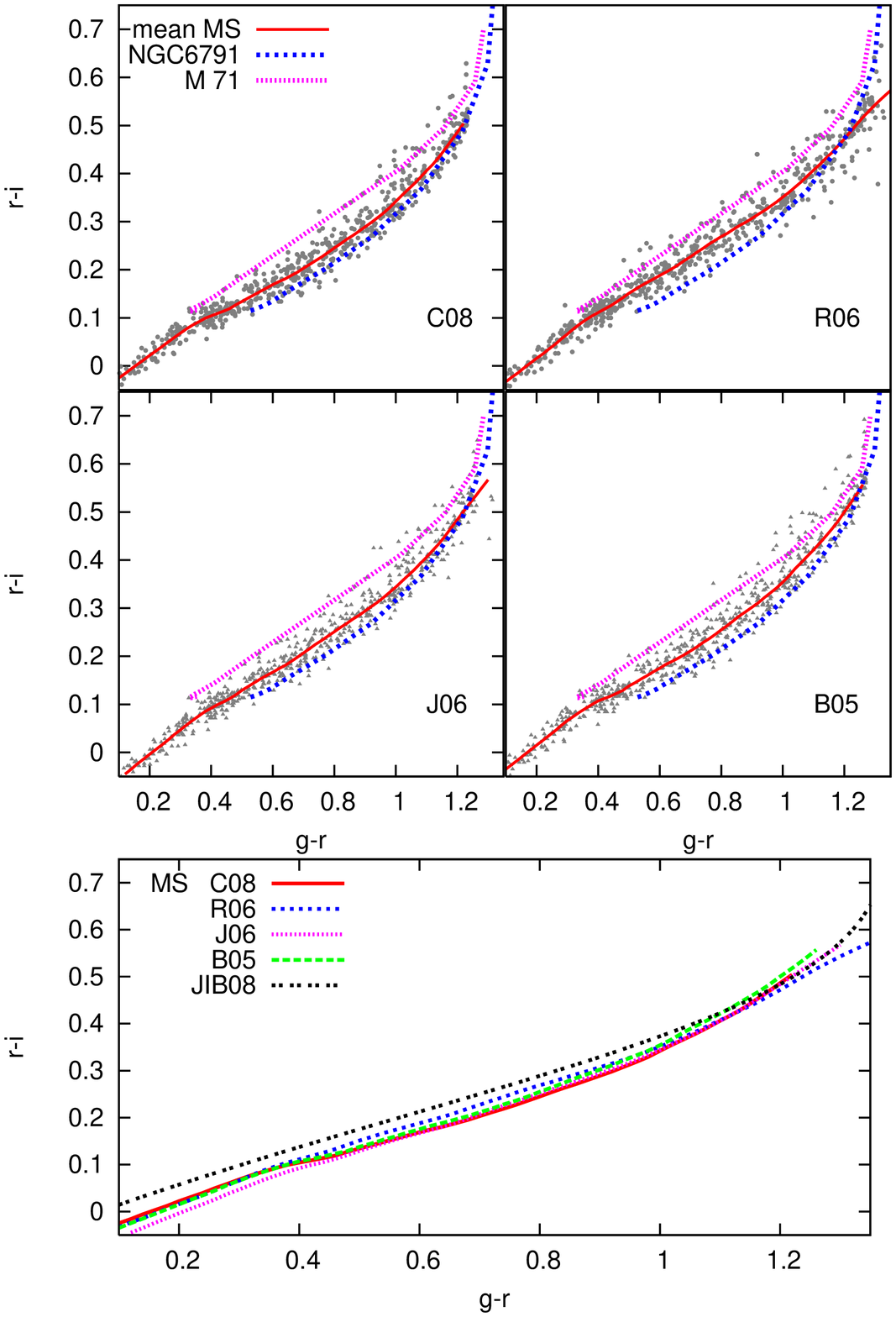}
\caption{The upper panels show similar plots as in Fig.\ \ref{fig-ms} 
but in ($g-r,r-i$). 
The lower panel shows a direct comparison of the mean MS
of the different transformations. The stellar locus of Juri\'{c} et al.
(\cite{Ju08} JIB08) is also shown.
\label{fig-2col2}
}
\end{figure}

Since 2-colour diagrams are independent of distance, they are a valuable tool to
investigate the intrinsic properties of the stellar populations. 
In Fig.\ \ref{fig-2col2} we
show the ($g-r,r-i$) diagrams for all four transformations 
compared to the ridge lines of the star clusters NGC 6791
and M71.  The transformations C08 and B05 are well bounded by the star cluster
fiducials, whereas the transformations R06 and J06, which use a ($V-R$)-term in
the colour transformation for $g-r$, show some outliers at the lower right.
Full lines are the mean MS, which are compared directly in the lower plot.
Additionally the stellar locus JIB08 of Juri\c{c} et al. (\cite{Ju08}) showing a
small offset to the lower metallicity regime.

\subsection{Local normalization}\label{sec-norm}

\begin{figure}
\includegraphics[height=8.5cm, angle=270]{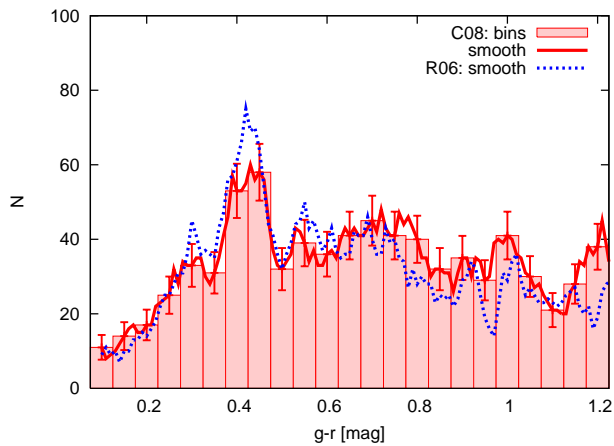}
\caption{ Local number density in the 25\,pc sphere in $g-r$ bins of 0.05\,mag.
The histogram and full line show the data with the
transformation C08 of Chonis \& Gascall (\cite{Ch08}). 
The errorbars show the $1\sigma$ noise error.
The dashed line is the result using the R06 transformation of 
Rodgers et al. (\cite{Ro06}) with the completeness corrections as discussed in
the text.
\label{fig-ngr}
}
\end{figure}

In order to determine the population at the MS as function of
colour $g-r$ a complete stellar sample is needed.
With C08 applied to the complete stellar sample of 761 stars in the 25\,pc
sphere we derive the number of stars per 0.05\,mag bin in $g-r$.
 The histogram in Fig.\ \ref{fig-ngr} shows the result for a selected position
 of the binning. Errorbars are $\sqrt{N}$ noise. The full line gives the
running mean, which show the sensitivity with respect to thechosen binning.

In order to investigate the sensitivity of the number density function on the
filter transformation, we derived $N(g-r)$ also for R06.
Since only 561 stars have all colours, we determine $M_\mathrm{g}$ and $g-r$ for the
other stars in an approximate way. For the 119 stars with $V-R_\mathrm{C}$ we estimate
$g-r$ from $g'-r'$ using an approximation from the stellar locus in the 2-colour
diagram ($g'-r',r'-i'$) for the 561 stars. The linear best fit of these
stars is
\bqn
r'-i' &=& (0.470\pm 0.004)(g'-r') - 0.086\pm 0.0035 \nonumber\\
  &&\quad \mbox{for}\quad  g'-r' \le 1.4
\eqn
leading to
\bqn
g-r &\approx& 1.044(g'-r')-0.021 \\
\eqn
instead of Eq. \ref{eq-grTu}.   For the remaining 106 stars we use
the transformation of C08 and correct $M_\mathrm{g}$ and $g-r$. For the
correction we determine a fourth order polynomial best fit of
$\Delta(g-r)$ (Fig.\ \ref{fig-dgr}) and add that to the
individual values from C08 of the stars. For testing the method, we applied the same
method to the subsample of 561 stars with all colours and compared the resulting
histograms with these of the R06 transformation. 
The differences are well below the uncertainty due to noise.

The dashed line shows the car-box average with the R06 transformation. There are
systematic differences (higher maximum, lower level at the red end), but they
are not significant due to the small number statistics.

The applications of the local number densities are twofold. Firstly they can be
used for disc models from star counts as the local normalization.
Secondly the local colour distribution can be used to understand the physical
properties of the stellar population in the solar neighbourhood.
 The steep increase at the blue end ($g-r<0.45$) is due to the
increasing lifetime of the stars with decreasing stellar mass until the age of
the disc is reached. This effect is reduced by the increasing scale height of
the stars with increasing mean age. The shape of the density function in the red
part  ($g-r>0.45$) is determined by a combination of the IMF and of the 
stellar mass-colour relation, which depends on the metallicity. 

\section{Isochrones}\label{sec-iso}

For an understanding of the physical properties of stellar populations a
comparison with theoretical isochrones is very useful.
Here we compare the data with two different sets of isochrones, where SDSS colours
are available and which are provided via a web interface.
The Padua isochrones \footnote{http://stev.oapd.inaf.it/cmd } were
based until recently on the evolutionary tracks of 
Girardi et al. (\cite{Gi02}). For the SDSS colours see Girardi et al.
\cite{Gi04}. Since 2008 new isochrones based on 
Marigo et al. (\cite{Ma08}) are provided via the web interface.

The Dartmouth Stellar Evolution Database 
\footnote{http://stellar.dartmouth.edu/~models/index.html} provides different
sets of isochrones in $B\,V\,R_\mathrm{C}\,I_\mathrm{C}$. 
We use the isochrones of VandenBerg \& Clem
(\cite{Va03}). In $u\, g\, r\, i\, z$ the isochrones are based on the Sixth
data release of SDSS (Adelman-McCarthy et al. \cite{Ad08}). 
The parameter grid in age is in steps of 0.5\,Gyr starting at an age of 1\,Gyr.
We use solar $\mathrm{[\alpha,Fe]}$ ratio and normal
Helium abundance in our investigation.

In Fig.\ \ref{fig-hrd} we compare the old and new Padua ZAMS 
(zero age main sequence = 1\,Myr isochrone) with metallicity 
$\mathrm{[Fe/H]}=+0.13$ in
in ($B-V,M_\mathrm{V}$) with the CMD of the full Hipparcos sample. There is  good
agreement in the blue and intermediate colour regime. 
At the red end the isochrones
are too blue/faint. The new isochrones (thin lines, labeled 2008) are
inferior to the classical isochrones (2007).
The Dartmouth isochrone of 1\,Gyr matches the MS very good down to the M dwarfs. 

\begin{figure}
\includegraphics[width=8.5cm, angle=0]{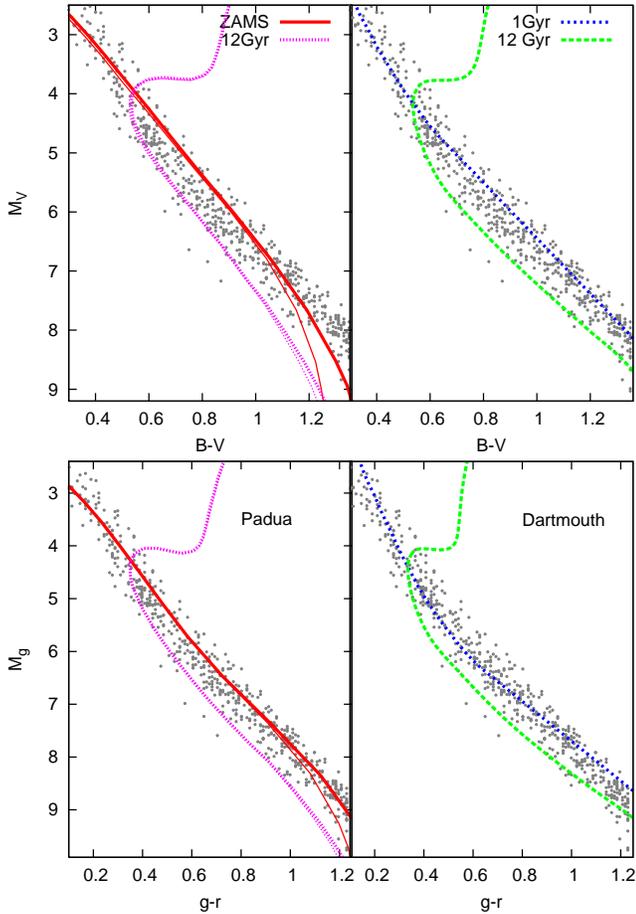}
\caption{CMDs in ($B-V,M_\mathrm{V}$) and in
($g-r,M_\mathrm{g}$) using the
transformation C08 for the same stars as in Fig. \ref{fig-ms}. 
The left panels show the Padua isochrones with [Fe/H]=+0.13 for the ZAMS and
[Fe/H]=-0.55 at age 12\,Gyr.  
The right panels show the Dartmouth isochrones with the same metallicities,
respectively.
\label{fig-hrd3}
}
\end{figure}

\begin{figure}
\includegraphics[width=8.5cm, angle=0]{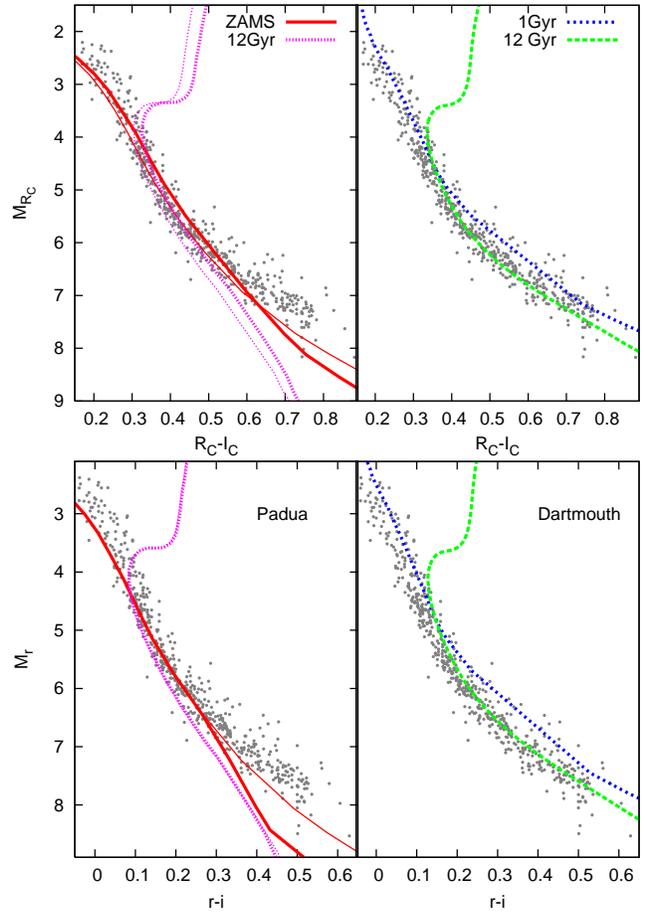}
\caption{CMDs in ($R_\mathrm{C}-I_\mathrm{C},M_{R_\mathrm{C}}$) and in
($r-i,M_\mathrm{r}$) with  stars and isochrones as in Fig. \ref{fig-hrd3}. 
\label{fig-hrd4}
}
\end{figure}

\begin{figure}
\includegraphics[width=8.5cm, angle=0]{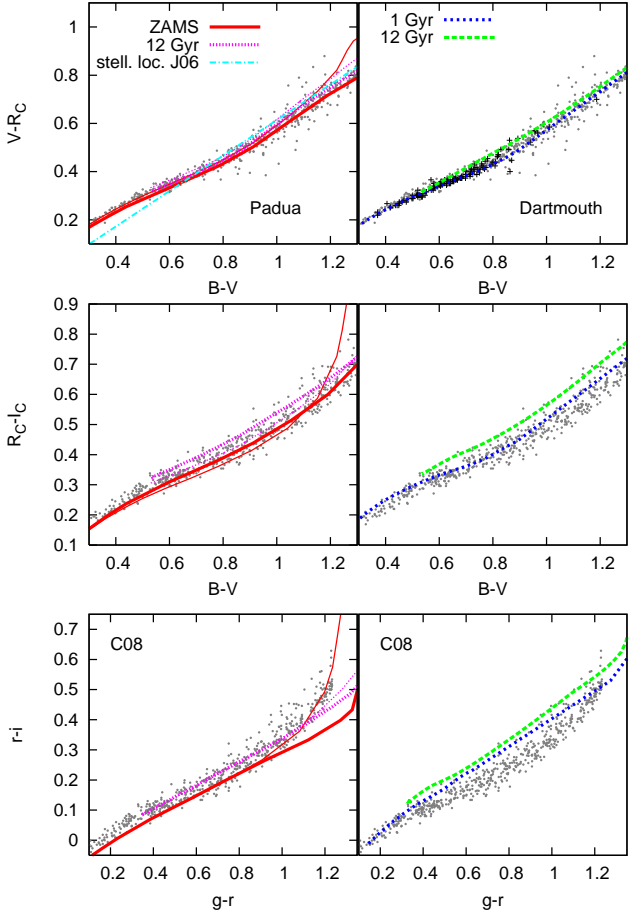}
\caption{2-colour diagrams of stars and isochrones as in Fig.\ \ref{fig-hrd3}.
In the upper left panel the thin line shows the stellar locus, 
where the set of equations of J06 are consistent with each other.
In the upper right panel the 119 stars with missing $R_\mathrm{C}-I_\mathrm{C}$ only are added
(crosses).
\label{fig-2col}
}
\end{figure}

A detailed comparison of the isochrones with the local stellar population
transformed with C08
is shown in Figs.\ \ref{fig-hrd3} - \ref{fig-2col}.
For Padua we show the ZAMS with  $\mathrm{[Fe/H]}=+0.13$ (as in Fig.\ \ref{fig-hrd}) and
the isochrone of age 12\,Gyr with $\mathrm{[Fe/H]}=-0.55$.
For Dartmouth we use the isochrones of age 1\,Gyr with  $\mathrm{[Fe/H]}=+0.13$
and  of age 12\,Gyr with $\mathrm{[Fe/H]}=-0.55$.
The upper panels show in 
($B-V,M_\mathrm{V}$) (Fig.\ \ref{fig-hrd3}),
($R_\mathrm{C}-I_\mathrm{C}, M_{R_\mathrm{C}}$) (Fig.\ \ref{fig-hrd4}),
($B-V,V-R_\mathrm{C}$) and ($B-V,R_\mathrm{C}-I_\mathrm{C}$) (Fig.\ \ref{fig-2col}),
where discrepancies in the Johnson-Cousins system are still present.
In ($B-V,V-R_\mathrm{C}$)  a number of stars scatter considerably around the
MS, which is not the case in ($B-V,R_\mathrm{C}-I_\mathrm{C}$). 
This  may be due to observational errors in $V-R_\mathrm{C}$.
In ($B-V,R_\mathrm{C}-I_\mathrm{C}$) the Dartmouth
isochrones are systematically offset to the red in ($R_\mathrm{C}-I_\mathrm{C}$). Also the
colour width due to the metallicity spread is too small.

The lower panels show similar diagrams in SDSS filters, namely
($g-r,M_\mathrm{g}$) (Fig.\ \ref{fig-hrd3}),
($r-i, M_\mathrm{r}$) (Fig.\ \ref{fig-hrd4}),
and ($g-r,r-i$) (Fig.\ \ref{fig-2col}).

In ($g-r,M_\mathrm{g}$) both sets of isochrones match the data with the C08 transformation
roughly. For G-K stars the Padua isochrones are better at the metal rich edge
(bright/red), whereas the Dartmouth isochrones are slightly better in the
transition regime from K to M dwarfs.
In ($r-i, M_\mathrm{r}$) the consistency is much worse. There are significant offsets of
of both sets of isochrones. The Padua isochrones are to faint/blue and the
Dartmouth isochrones are too bright/red.

In the two-colour diagrams the picture is different. Here the new Padua 
ZAMS fit better than the old one, but the colour spread along the MS cannot be
reproduced. For the Dartmouth isochrones there is a large offset at the young
and metal rich end.

\begin{figure}
\includegraphics[width=8.5cm, angle=0]{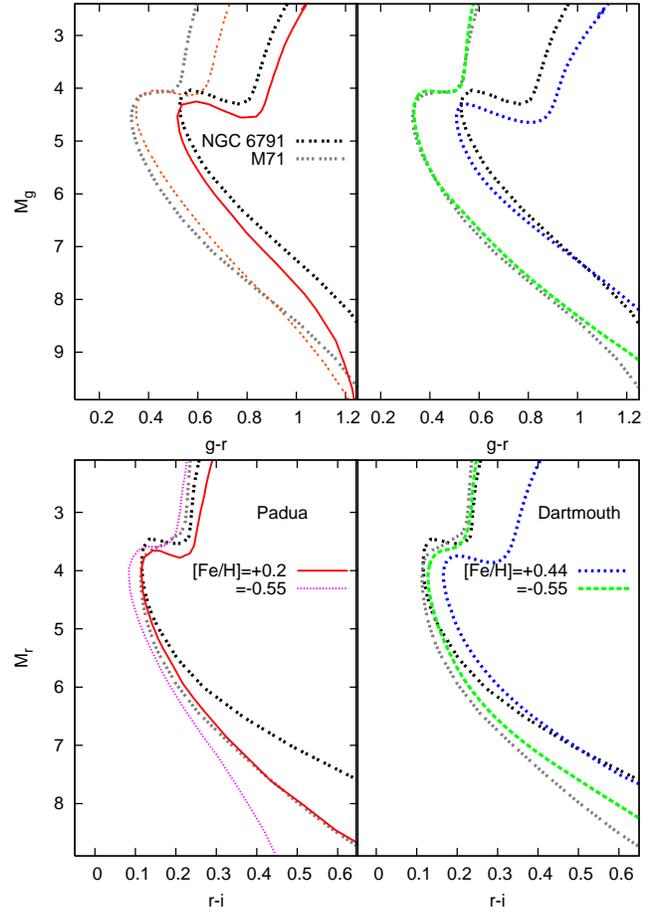}
\caption{ Comparison of the fiducial star clusters NGC 6791 and M71 with the
corresponding isochrones from 
Padua (left panels: 8\,Gyr, [Fe/H]=+0.2 and 12\,Gyr, [Fe/H]=-0.55)
and Dartmouth (right panels: 8\,Gyr, [Fe/H]=+0.44 and 12\,Gyr, [Fe/H]=-0.55).
\label{fig-isohrd}
}
\end{figure}

\begin{figure}
\includegraphics[height=8.5cm, angle=270]{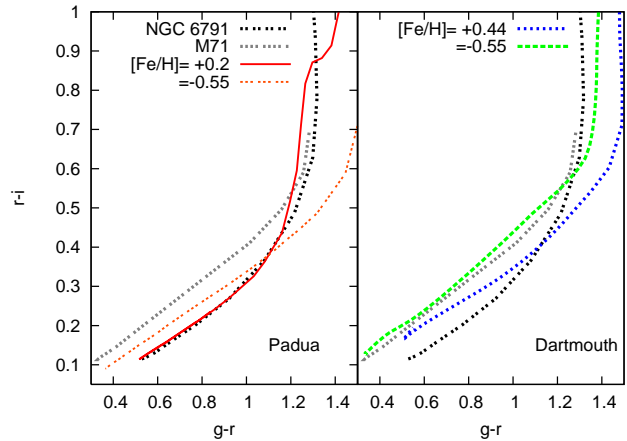}
\caption{ Same cluster ridge lines and isochrones as in Fig.\ \ref{fig-isohrd}
in the 2-colour diagrams.
\label{fig-isogri}
}
\end{figure}

Finally we compare the ridge lines for the star clusters  M71 and NGC 6791
of Clem et al. (\cite{Cl08}) with isochrones. We M71 we use an age of
12\,Gyr and $\mathrm{[Fe/H]}=-0.55$. For the old metal rich cluster NGC 6791 we
use an age of 8\,Gyr and choose the closest available metallicity, which is 
$\mathrm{[Fe/H]}=+0.2$ for Padua and $\mathrm{[Fe/H]}=+0.44$ for Dartmouth. The
results are shown in Figs.\ \ref{fig-isohrd} and \ref{fig-isogri}. 
Generally there is a better agreement of the Dartmouth isochrones with the
ridge lines of the star clusters in ($g-r,M_\mathrm{g}$). 
The offsets of both sets of isochrones in ($r-i,M_\mathrm{r}$) are much larger
with a similar systematics as for the local stellar sample. In the two-colour
diagrams is becomes again obvious that the variation in colour due to
metallicity spread is too small for F to K stars and too large in the M star
regime.

\section{Summary\label{sec-disc}}

We transformed the MS of F to K dwarfs in the solar neighbourhood
from Johnson Cousins $B\,V\,R_\mathrm{C}\,I_\mathrm{C}$ to SDSS colours $g\,r\,i$ using the
transformations of different authors. 
We used the complete sample of stars in a 25\,pc sphere around the Sun 
from the Catalogue of Nearby Stars (CNS4) with Hipparcos distances to determine
the mean MS properties in ($g-r,M_\mathrm{g}$) with the transformation C08 
of Chonis \& Gascall (\cite{Ch08}). 
With the subsample, where all colours are
available, we determined the mean MS in
the CMDs in ($g-r,M_\mathrm{g}$) and ($r-i,M_\mathrm{r}$) and the two-colour diagram
($g-r,r-i$). 
The systematic differences between the transformations of different authors
in $M_\mathrm{g}$, $M_\mathrm{r}$, $g-r$ and $r-i$ are below 0.1\,mag.
In the cases, where $V-R_\mathrm{C}$ is used in the transformations a considerable
individual scatter up to 0.3\,mag are introduced. This was not expected, when
using more colour-terms in the transformations. 
The position of the mean MS in the CMDs is  very sensitive to systematic
deviations in the transformations. The differences in $M_\mathrm{g}(g-r)$ are up to
0.5\,mag and in $M_\mathrm{r}(r-i)$ up to 0.3\,mag, which is significant for photometric
distance estimations. Also the 'bright' MS of Juri\'{c} et al. (\cite{Ju08})
is consistent with the local stellar sample. 

The distribution of the solar neighbourhood stars in the CMDs and the
two-colour diagram agree reasonably well 
with fiducial colour magnitude sequences of star clusters observed in
$g'\,r'\,i'$. The best agreement occurs with the transformation C08 
of Chonis \& Gascall (\cite{Ch08}).

The comparison with theoretical isochrones, especially in 
($r-i,M_\mathrm{r}$) and ($g-r,r-i$), is less  satisfying. The systematic differences in
shape and location require further calibration by observations in the SDSS
filter system.

The solar neighbourhood provides a good testcase for the interpretation of
mixed stellar populations. The results presented here can be used for detailed
models of the Milky Way disc based on star counts, 
when extrapolated to the bright end in apparent magnitude.
For this application we provide the local stellar number densities as function
of colour $g-r$.

\end{document}